\documentclass{jps-cp}
\usepackage{txfonts} 
\usepackage{amsmath,amssymb,bm}
\usepackage{graphicx}
\usepackage{color}
\title{Temperature evolution of magnetic phases near the thickness-dependent metal-insulator transition in La$_{1-x}$Sr$_{x}$MnO$_3$ thin films observed by XMCD}

\author{Goro \textsc{Shibata}$^{1}$, Kohei \textsc{Yoshimatsu}$^2$, Enju \textsc{Sakai}$^2$, Keisuke \textsc{Ishigami}$^1$, Shoya \textsc{Sakamoto}$^1$, 
Yosuke \textsc{Nonaka}$^1$, Fan-Hsiu \textsc{Chang}$^3$, Hong-Ji \textsc{Lin}$^3$, Di-Jing \textsc{Huang}$^3$, Chien-Te \textsc{Chen}$^3$, Hiroshi \textsc{Kumigashira}$^2$, 
and Atsushi \textsc{Fujimori}$^1$}

\inst{
$^1$Department of Physics, The University of Tokyo, Tokyo 113-0033, Japan \\
$^2$Institute of Materials Structure Science, High Energy Accelerator Research
Organization (KEK), Ibaraki 305-0801, Japan \\
$^3$National Synchrotron Radiation Research Center, Hsinchu 30076, Taiwan \\
}

\email{shibata@rs.tus.ac.jp}

\abst{
Perovskite-type manganites, which are well-known for their 
intriguing physical properties such as colossal magnetoresistance (CMR) 
and half metalicity, have been considered as candidate materials for spintronics. 
However, their ferromagnetic (FM) properties are often suppressed 
in thin films when the thickness is reduced down to several monolayers (MLs). 
In order to investigate how the magnetic phases evolve near the paramagnetic (PM)-to-FM phase transition boundary, 
we have performed temperature-dependent x-ray magnetic circular dichroism (XMCD) 
experiments on a La$_{1-x}$Sr$_{x}$MnO$_3$ (LSMO, $x=0.4$) thin film, 
whose thickness (8 ML) is close to the boundary between the FM-metallic and the PM-insulating phases. 
By utilizing the element-selectiveness of XMCD, 
we have quantitatively estimated the fractions of the 
PM and superparamagnetic (SPM) phases as well as the FM one 
as a function of temperature. 
The results can be reasonably described based on a microscopic phase-separation model. 
}

\kword{manganites, thin films, x-ray magnetic circular dichroism, phase separation, colossal magnetoresistance} 

\begin{document}
\maketitle

\section{Introduction}

Perovskite-type manganese oxides have drawn much attention for decades 
because of their wide varieties of electric and magnetic properties \cite{IFT}. 
They exhibit complex phase diagrams as a function of temperature and hole concentration 
due to mutual coupling of spin, charge, orbital, and lattice degrees of freedom. 
One of the most notable phenomena of the perovskite-type manganites is
colossal magnetoresistance (CMR) \cite{LSMO_CMR_JPSJ94, Ramirez_CMR_JPCM97, Tokura_CMR_JMMM99, Tokura_CMR_06}, 
which stands for the large reduction of electric resistivity upon applying 
external magnetic field. 
The CMR effect has been understood as the phase competition 
between the Mott-insulating phase and the ferromagnetic-metallic (FM-M) phase 
due to double-exchange (DE) interaction: 
The external magnetic field stabilizes the FM-M phase 
by aligning the directions of Mn spins, 
leading to the reduction of the resistivity. 
Indeed, phase separation 
in a microscopic scale has been observed 
by several microscopy observations \cite{phasesep_Uehara, phasesep_Fath, phasesep_LPCMO}. 

The perovskite-type manganite La$_{1-x}$Sr$_{x}$MnO$_3$ (LSMO) 
has been considered as a promising material for spintronics 
because it has the highest Curie temperature ($T_\text{C}$) 
among the manganites ($T_\text{C} \sim 360\ \text{K}$ 
at the optimal doping levels of $x = 0.2$--$0.4$) \cite{IFT, LSMO_Urushibara, LSMO_phasediag} 
and almost full spin polarization at the Fermi level (half metalicity). 
For this reason, 
growth of high-quality LSMO thin films has been attempted by various groups. 
It has been known, however, that the physical properties of LSMO thin films 
are strongly influenced by external conditions, 
such as epitaxial strain from the substrate, surface and interfacial effects, 
and dimensional effects. 
These effects often lead to suppression or loss of the FM-M properties \cite{Konishi}. 
In particular, both the saturation magnetization and $T_\text{C}$ are significantly reduced 
when the film thickness decreases down to several monolayers (MLs) 
\cite{LSMO_Huijben, LSMO_Yoshimatsu, LSMO_Shibata}, 
which is a drawback to the realization of 
nanoscale spintronics devices. 


In order to elucidate the origin of the loss of ferromagnetism in ultrathin LSMO films, 
it is necessary to clarify how the magnetic phases evolve 
in the vicinity of the FM-paramagnetic (PM) phase boundaries. 
In the present study, we have investigated the magnetism of LSMO ($x = 0.4$) ultrathin films with reduced thickness (and therefore reduced $T_\text{C}$) 
using x-ray absorption spectroscopy (XAS) and x-ray magnetic circular dichroism (XMCD) 
as functions of temperature. 
Owing to the element-selectiveness of XMCD spectroscopy, 
one can probe the intrinsic magnetic properties of thin films 
without contributions from the diamagnetism/paramagnetism of the substrate. 
This makes it possible to quantitatively estimate various aspects of magnetic properties 
such as intrinsic paramagnetism, average moment in the superparamagnetic (SPM) phase, 
and the volume fraction of each phase 
\cite{Takeda_GaMnAs, LSMO_Shibata, InFeAs_Sakamoto}. 
We have shown that the evolution of the magnetism near $T_\text{C}$ 
in a thin film of 8 ML, which is the critical thickness  between the FM-M and PM phases, 
can be understood 
as a phase mixture model of FM, SPM, and PM phases, 
similar to those observed in 
bulk manganite crystals \cite{phasesep_Uehara, phasesep_Fath, phasesep_LPCMO}. 

\section{Experiments}

LSMO thin films ($x = 0.4$) were grown on SrTiO$_3$ (STO) (001) substrates by the laser molecular beam epitaxy method. 
During the growth, the thickness of the LSMO film was monitored by the intensity oscillation of reflection high-energy electron-diffraction (RHEED) signals, 
and films of 8 ML ($\sim 3\ \text{nm}$) thickness were grown. 
After the deposition of LSMO, 1 ML of La$_{0.6}$Sr$_{0.4}$TiO$_3$ (LSTO) and subsequently 2 ML of STO were deposited as capping layers. 
The LSTO layer was inserted in order to prevent hole doping at the LSMO/STO interface \cite{LSMO_Yoshimatsu}. 
Prior to the XMCD measurement, the samples were annealed in oxygen atmosphere of 1 atm at 400 $^{\circ}\text{C}$ for 45 minutes 
in order to eliminate oxygen vacancies. 
The surface morphology of the films was checked using atomic force microscopy, and atomically-flat step-and-terrace structures 
were observed. 
The crystal structure was characterized by four-circle x-ray diffraction and the coherent growth of the LSMO layers on the STO substrates was confirmed. 
The XAS and XMCD spectra were measured at BL-11A of 
Taiwan Light Source, National Synchrotron Radiation Research Center (NSRRC), 
using the total electron-yield mode. 
The experimental geometry is shown as the inset of Fig.\ \ref{spectra}(b). 
The x-ray incident angle was 30$^\circ$ from the sample surface. 
The degree of the circular polarization ($P_c$) was 
$\sim 60\%$ \cite{Takahashi_SFMO}. 
The magnetic field ($\bm{H}$) was applied parallel to the LSMO thin film (along the [100] direction 
in the pseudocubic notation), 
which is the easy magnetization direction of the sample. 
The maximum magnetic field was 1 T.

\section{Results and Discussion}

\begin{figure}[tbh]
\centering
\includegraphics[width=6.5cm]{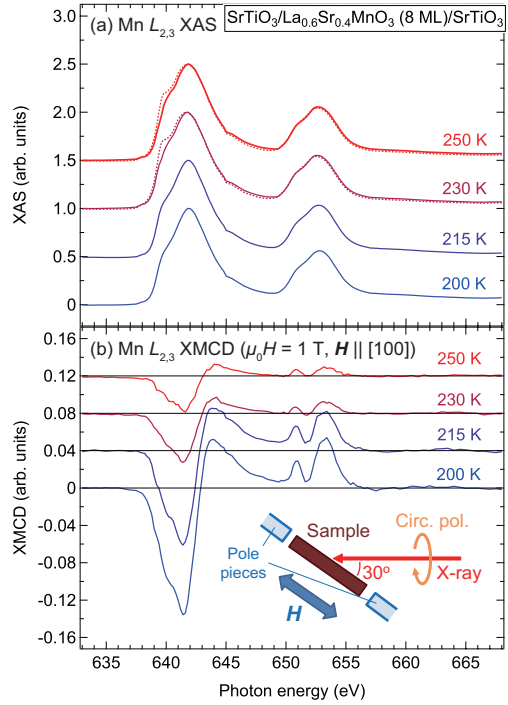}
\caption{
Mn $L_{2,3}$ ($2p_{1/2}, 2p_{3/2} \rightarrow 3d$) 
x-ray absorption spectroscopy (XAS) (a) and x-ray magnetic circular dichroism (XMCD) (b) spectra 
of the 8-ML-thick La$_{1-x}$Sr$_{x}$MnO$_3$ (LSMO, $x=0.4$) thin film 
measured at different temperatures. 
In panel (a), dotted lines are the raw XAS spectra 
and solid lines are the spectra after subtracting the extrinsic Mn$^{2+}$ signals 
(see text). 
Inset in panel (b) shows the experimental geometry.
}
\label{spectra}
\end{figure}

Figure\ \ref{spectra}(a) shows the Mn $L_{2,3}$-edge ($2p_{1/2}, 2p_{3/2} \rightarrow 3d$) 
XAS spectra of the 8-ML-thick LSMO thin film 
measured at various temperatures around $T_\text{C}$. 
The spectral line shapes are similar to those reported in previous studies 
\cite{LSMO_Koide, LSMO_Shibata}, 
except for the weak shoulder structure around $h\nu=640\ \text{eV}$ 
which appears in the spectra for $T=250\ \text{K}$ and $T=230\ \text{K}$. 
This structure is characteristic of Mn$^{2+}$ ions,
which may originate from oxygen vacancies at the film surface. 
It has been pointed out that such extrinsic Mn$^{2+}$ signals are often observed 
in the XAS spectra of manganite thin films \cite{Mn2plus}.
We have therefore subtracted these extrinsic Mn$^{2+}$ spectra 
from the raw XAS spectra for $T=250\ \text{K}$ and $T=230\ \text{K}$ 
by using a reference XAS spectra. 
The XAS spectra after the subtraction are shown in Fig.\ \ref{spectra}(a)
by the solid lines. 
The intensity of the Mn$^{2+}$ signals relative to the entire XAS intensity 
has been estimated to be $\sim 7 \pm 2\%$
for both the spectra. 
After the subtraction of the Mn$^{2+}$ signals, 
all the spectra show almost the same line shapes 
regardless of the temperature. 
This suggests that the intrinsic valence states of Mn ions 
are not significantly affected by 
whether the films are in the metallic or insulating states. 

Figure\ \ref{spectra}(b) shows the Mn $L_{2,3}$-edge XMCD spectra corresponding to Fig.\ \ref{spectra}(a). 
Here, the XMCD spectra have been normalized to the peak height of the helicity-averaged XAS spectra 
at the Mn $L_3$ ($2p_{3/2}$) edge. 
With decreasing temperature, the XMCD intensity gradually increases. 
We note that all the spectra show essentially the same line shapes, 
indicating that the extrinsic Mn$^{2+}$ ions do not contribute to the magnetism. 

\begin{figure}[tbh]
\centering
\includegraphics[width=7cm]{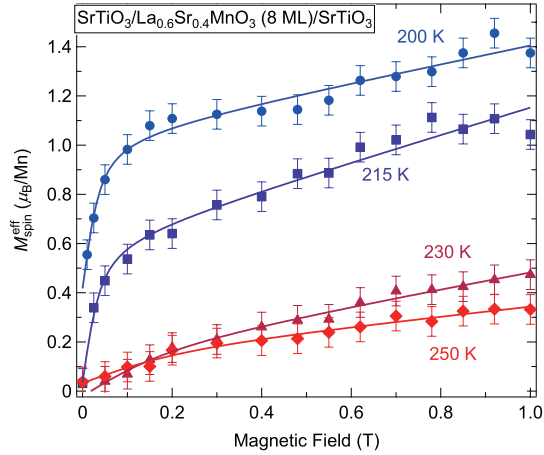}
\caption{
Magnetization curves of the 8-ML-thick LSMO thin film at various temperatures 
around $T_\text{C}\ (\sim 230\ \text{K})$ 
deduced from the XMCD measurements. 
Symbols: experimental data, Solid lines: fitted curves 
using the sum of a Langevin function, a linear function, 
and a constant, representing the SPM, PM, and FM components, respectively.
}
\label{MH}
\end{figure}

Figure\ \ref{MH} shows the magnetic-field ($H$) dependence of the 
spin magnetic moment ($M_{\text{spin}}$) of Mn 
deduced by using the XMCD spin sum rule \cite{SpinSum}. 
Here, the orbital magnetic moment and the magnetic dipole term \cite{SpinSum, TXMCD_Stohr} 
have been neglected because $M_{\text{spin}}$ has the dominant ($>90\%$) 
contribution to the total magnetic moment. 
Detailed procedure for the sum rule analysis 
is described in Refs.\ \cite{LSMO_Shibata} and \cite{LSMO_Koide}. 
Since the magnetic-field direction was 30$^{\circ}$ off the incident x rays 
and the degree of circular polarization $P_c$ was $\sim 60\%$, 
the magnetic moments deduced from the XMCD sum rule 
are divided by the factor $0.6\cos 30^\circ$ 
in order to deduce the magnetic moment. 
According to the magnetization measurement 
using a superconducting quantum interference device (SQUID) magnetometer \cite{LSMO_Yoshimatsu}, 
the magnetization of the 8-ML-thick LSMO thin film is estimated to be 
$\sim 30\ \text{emu/cm}^3$ ($\sim 0.2\ \mu_\text{B}\text{/Mn}$) 
at $T=200\ \text{K}$ and $\mu_0 H = 0.05\ \text{T}$. 
The $M_\text{spin}$ values in Fig.\ \ref{MH} are larger than it 
($\sim 0.86\ \mu_\text{B}\text{/Mn}$ at $T=200\ \text{K}$ and $\mu_0 H = 0.05\ \text{T}$), 
presumably due to systematic errors in estimating the sample size 
and the diamagnetic contribution in the SQUID magnetization measurement, 
and/or the errors in $P_c$ and the number of holes in the Mn 3$d$ bands ($n_h$) 
used in the XMCD sum rule 
\cite{SpinSum}. 
In the same magnetization measurement \cite{LSMO_Yoshimatsu},
the Curie temperature of the 8-ML-thick LSMO thin film is 
estimated to be $T_\text{C} \sim 230\ \text{K}$ 
(which is lower than that of the bulk due to the thickness effect \cite{LSMO_Yoshimatsu}). 
At higher temperatures, the remanent magnetization is almost zero 
and the shape of the magnetization curves is close to that of SPM materials (Langevin function). 
With decreasing temperature, the remanent magnetization becomes finite 
but the magnetization still increases almost linearly in $H$ 
at higher magnetic fields. 
In order to see this tendency more quantitatively, 
we have fitted the data to the sum of a Langevin function, a linear function, 
and a constant, representing the SPM, PM, and FM components, 
namely, 
\begin{equation}
M_{\text{spin}}\ (\mu_B/\text{Mn}) = 3.6 P_\text{SPM} \left[ \coth \left( \frac{\mu \mu_0 H}{k_B T} \right) - \left( \frac{k_B T}{\mu \mu_0 H} \right) \right]  + P_\text{PM} \chi \mu_0 H + 3.6P_\text{FM}, \label{fiteq} \\
\end{equation}
\begin{equation}
P_\text{SPM} + P_\text{PM} + P_\text{FM} = 1.
\end{equation}
where $P_\text{SPM}$, $P_\text{PM}$, and $P_\text{FM}$ are the fractions of 
the Mn atoms in the SPM, FM, and PM phases, respectively, 
$\chi$ is the PM susceptibility per Mn atom, 
and $\mu$ is the average magnetic moment of a single SPM cluster. 
Here, we assume that the magnetic moment of a single Mn atom in the FM and SPM phases 
is equal to the average magnetic moment of La$_{0.6}$Sr$_{0.4}$MnO$_3$ 
($3.6\ \mu_B/\text{Mn}$). 
The fitted magnetization curves are shown in Fig.\ \ref{MH}. 
One can see that the experimental results are fitted by Eq.\ (\ref{fiteq}) reasonably well, 
demonstrating that the magnetic state of the film can be described as 
a mixture of the FM, SPM, and PM phases. 

\begin{figure}[tbh]
\centering
\includegraphics[width=14cm]{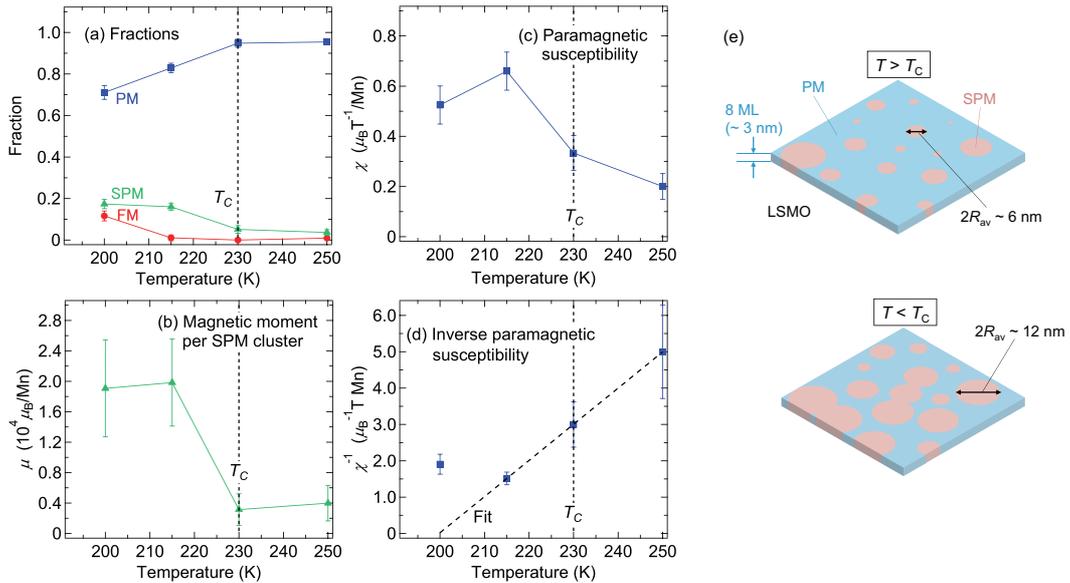}
\caption{
Temperature dependencies of (a) fractions of the PM, SPM, and FM magnetic phases 
($P_\text{PM}$, $P_\text{SPM}$, and $P_\text{FM}$), 
(b) average magnetic moment per SPM cluster ($\mu$), 
(c) PM susceptibility per Mn atom ($\chi$), 
and (d) inverse susceptibility $\chi^{-1}$.
$T_\text{C}$ indicated in the figure (230 K) is the value deduced from the magnetization measurements 
using a superconducting quantum interference device (SQUID) magnetometer 
\cite{LSMO_Yoshimatsu}. 
Black dashed line represents a linear fit based on the Curie-Weiss law. 
(e) Schematic description of the SPM domain sizes deduced from (b).
}
\label{decompo}
\end{figure}

Figure \ref{decompo} shows the temperature dependencies of the deduced fit parameters 
in Eq.\ (\ref{fiteq}): the fraction of each magnetic phase 
($P_\text{PM}$, $P_\text{SPM}$, and $P_\text{FM}$), 
the magnetic moment of a single SPM cluster ($\mu$), 
the PM susceptibility ($\chi$), and the inverse susceptibility. 
As shown in Fig.\ \ref{decompo}(a), the PM phase is dominant above $T_\text{C}\ (\sim 230\ \text{K})$, 
although some amount of SPM clusters do exist in the same temperature range. 
With decreasing temperature below $T_\text{C}$, both the fraction of the SPM phase 
and the average SPM cluster size rapidly increase, as shown in Figs.\ \ref{decompo}(a) 
and \ref{decompo}(b), whereas the fraction of the FM phase start to increase 
at a lower temperature. 
This is consistent with the microscopic phase-separation model in that 
small SPM clusters are formed prior to the PM-to-FM transition 
near $T_\text{C}$. 
From Fig.\ \ref{decompo}(b), the average radius of the SPM clusters is estimated to be 
$\sim 3\ \text{nm}$ at higher temperatures and $\sim 6\ \text{nm}$ at lower temperatures, 
under the assumption that the height of the cluster is equal to the film thickness 
($=8\ \text{ML} \sim 3\ \text{nm}$), as described in Fig.\ \ref{decompo}(e). 
We also note that more than half of the Mn atoms 
still remain in the PM phase even at $T=200\ \text{K}$. 

Figures \ref{decompo}(c) and \ref{decompo}(d) show the PM susceptibility per atom 
and its inverse, respectively. As shown in Fig.\ \ref{decompo}(d), 
the inverse susceptibility can be fitted by a linear function 
for the data of $T \geq 215\ \text{K}$. 
According to the Curie-Weiss law, the PM susceptibility $\chi$ above $T_\text{C}$ is given by 
\begin{equation}
\chi = \frac{C}{T-\Theta}, 
\end{equation}
where $C= g^2 \mu_B S(S+1)/(3k_B)$ is the Curie constant with atomic spin $S$ 
and $\Theta$ is the Weiss temperature. 
Assuming that the Curie constant of LSMO ($x=0.4$) is given by the weighted average of 
$C$'s of Mn$^{3+}$ ($S=2$) and Mn$^{4+}$ ($S=3/2$) with the ratio of 6:4, 
the average $C$ is calculated to be $\simeq 4.57\ (\mu_B \text{T}^{-1}/\text{Mn})$. 
Then, the Weiss temperature of the present 8-ML-thick LSMO thin film is estimated to be 
$\Theta = 200\ \text{K}$ from the linear fit in Fig.\ \ref{decompo}(d). 
This is smaller than the Curie temperature $T_\text{C}$ deduced from the 
magnetization measurement \cite{LSMO_Yoshimatsu}. 
This may imply that the ferromagnetic spin correlation in the PM phase 
is weaker than those in the FM and SPM phases. 
The drop of $\chi$ at $T=200\ \text{K}$ may indicate that 
the average ferromagnetic spin correlation in the remaining PM phase 
further decreases as the FM phase expands. 

\section{Summary}
We have studied the evolution of magnetic phases in the 8-ML-thick LSMO thin film 
across the PM-to-FM transition via XMCD. 
The magnetization curves can be accounted for as sums of the 
FM, SPM, and PM components, supporting the phase-separation model. 
It is found that more than half of the Mn atoms remain in the PM phase 
even below the $T_\text{C}$. 
The volume fraction of the SPM phase increases 
across the macroscopic $T_\text{C}$ of the film (230 K), 
with the average radius of the SPM cluster 
also increasing from $\sim 3\ \text{nm}$ to $\sim 6\ \text{nm}$. 
The Weiss temperature in the PM phase has been estimated to be $\Theta = 200\ \text{K}$, 
which is lower than the macroscopic $T_\text{C}$. 
This may imply the weakening of the ferromagnetic correlation in the PM phase. 

\section*{Acknowledgments}
This work was supported by a Grant-in-Aid for Scientific Research from 
the JSPS (Project No. 22224005).
The experiment was done under the approval of 
NSRRC 
(proposal No.\ 2014-2-111). 


\begin{thebibliography}{9}
\bibitem{jpsj} The abbreviation for the Journal of the Physical Society of Japan should be ``J. Phys. Soc. Jpn." in the reference list.
\end{thebibliography}
\end{document}